\documentclass{PoS}
\usepackage{graphicx}
\usepackage{amssymb}
\usepackage{amsmath}
\usepackage{bm}
\usepackage{slashed}
\usepackage{datetime}
\usepackage{mciteplus}
\usepackage{multirow}
\usepackage{color}
\usepackage[utf8]{inputenc}
\usepackage[normalem]{ulem}
\usepackage{float} 
\usepackage[caption=false]{subfig} 

\newcommand{\smarrow}{\mbox{\raisebox{-4.5pt}[0pt][0pt]{$\hspace{-1pt} 
		\vec{\phantom{v}}$}}}
\newcommand{\hermes}{\textsc{Hermes}}
\newcommand{\compass}{\textsc{Compass}}

\newcommand{\T}{\perp}
\newcommand{\Tperp}{T}
\newcommand{\bT}{\xi_T}
\newcommand{\bb}{\xi}

\title{A first determination of the unpolarized quark TMDs from a global analysis}

\ShortTitle{A first determination of the unpolarized quark TMDs from a global analysis}

\author{\speaker{Cristian Pisano}\\
        Dipartimento di Fisica, Universit\`a di Pavia and INFN, Sezione di Pavia \\
        Via Bassi 6, I-27100 Pavia, Italy\\
        Dipartimento di Fisica, Universit\`a di Cagliari and INFN, Sezione di Cagliari \\
        Cittadella Universitaria, I-09042 Monserrato (CA), Italy\\
        E-mail: \email{cristian.pisano@ca.infn.it}}

\author{Alessandro Bacchetta\\
         Dipartimento di Fisica, Universit\`a di Pavia and INFN, Sezione di Pavia \\
        Via Bassi 6, I-27100 Pavia, Italy\\
        E-mail: \email{alessandro.bacchetta@unipv.it}}

\author{Filippo Delcarro\\
        Dipartimento di Fisica, Universit\`a di Pavia and INFN, Sezione di Pavia \\
        Via Bassi 6, I-27100 Pavia, Italy\\
        E-mail: \email{filippo.delcarro@pv.infn.it}}

\author{Marco Radici\\
        INFN, Sezione di Pavia \\
        Via Bassi 6, I-27100 Pavia, Italy\\
        E-mail: \email{marco.radici@pv.infn.it}}

\author{Andrea Signori\\
        Theory Center, Thomas Jefferson National Accelerator Facility\\
        12000 Jefferson Avenue, Newport News, VA 23606, USA\\
        E-mail: \email{asignori@jlab.org}}

\abstract{Transverse momentum dependent distribution and fragmentation functions of unpolarized quarks inside unpolarized protons are extracted, for the first time, through a simultaneous analysis of semi-inclusive deep-inelastic scattering, Drell-Yan and $Z$ boson hadroproduction processes. This study is performed at leading order in perturbative QCD, with energy scale evolution at the next-to-leading logarithmic accuracy.  Moreover, some specific choices are made to deal with low scale evolution around 1 GeV$^2$. Since only data in the low transverse momentum region are considered, no matching to fixed-order calculations at high transverse momentum is needed.}

\FullConference{QCD Evolution 2017\\
                 22-26 May, 2017\\
                 Jefferson Lab, Newport News, VA - USA}

\begin{document}

\section{Introduction}

Transverse momentum dependent parton distribution and fragmentation functions (TMDs) are fundamental objects that encode information on the intrinsic motion of quarks and gluons inside hadrons. As such, they provide a three-dimensional picture of hadrons in momentum space. In analogy to the more common {\em collinear}  parton distribution and fragmentation functions (integrated over transverse momentum), TMDs are not purely perturbative quantities, which could be fully derivable from first principles. Only their dependence on the energy scale can be calculated in perturbative QCD. However, in contrast to DGLAP equations which govern the evolution of collinear distributions,  the TMD evolution kernels contain a nonperturbative part as well. Hence, the complete determination of TMDs is not at all trivial, and requires a continuos interplay between theory and experiment. 

In this contribution to the proceedings, based on Ref.~\cite{Bacchetta:2017gcc} to which we refer for further details, we present a first attempt to extract unpolarized quark TMDs through a global fit. To this aim, we consider only those processes for which factorization has been established~\cite{Collins:2011zzd} and data are already available, namely semi-inclusive deep-inelastic scattering (SIDIS), Drell-Yan (DY) and $Z$ boson production in proton-proton collisions. We note that electron-positron annihilation into two hadrons is an important reaction that is, however, currently missing. Its future analysis could allow for an independent study of transverse momentum dependent fragmentation functions~\cite{Bacchetta:2015ora}.

Since the data included in the present study come from many experiments, carried out at different energies, our results not only shed light on the partonic transverse momentum dependence, but they represent a test of TMD evolution over a large energy range. Furthermore,  at the same time, we are able to probe the universality properties of the TMDs among different processes.

This contribution is organized as follows. In section~\ref{sec:frame}, we briefly outline the adopted TMD framework.  In section~\ref{sec:evol} we clarify our approximations in the phenomenological implementation of TMD evolution equations and describe the nonperturbative input of the fit. Our results are presented and discussed in section~\ref{sec:res}. A summary and suggestions for future possible improvements are provided in section~\ref{sec:conc}.

\section{Theoretical Framework}
\label{sec:frame}
The cross section  $\sigma_N^h$ and transverse structure function  $F_{UU,T}$ for the SIDIS process $\ell(l) + N(P) \to \ell(l') + h(P_h) + X$ are commonly expressed in terms of the variables
\begin{align}
  \label{e:xyz}
x &= \frac{Q^2}{2\,P\cdot q}\,,
&
y &= \frac{P \cdot q}{P \cdot l}\,,
&
z &= \frac{P \cdot P_h}{P\cdot q}\,,
\end{align}
with $Q^2=-q^2 = -(l-l^\prime)^2$. If we denote the corresponding quantities for the fully inclusive DIS process $\ell(l) + N(P) \to \ell(l')  + X$ by $\sigma_{\text{DIS}}$ and $F_T$, it is possible to define the hadron multiplicity as follows, 
\begin{equation}
m_N^h (x,z,|\bm{P}_{h\Tperp}|, Q^2) = \frac{d \sigma_N^h / ( dx \, dz \,d|\bm{P}_{h\Tperp}|\, dQ^2) }
                                                                   {d\sigma_{\text{DIS}} / ( dx \,dQ^2 ) }\approx \frac{2 \pi\,|\bm{P}_{h\Tperp}| F_{UU ,T}(x,z,\bm{P}_{h\Tperp}^2, Q^2)}{F_{T}(x,Q^2) } \,.
\label{e:multiplicity}
\end{equation}
We point out that the last approximation in (\ref{e:multiplicity}) is only valid in the kinematic region under study, {\it i.e}  $\bm{P}_{hT}^2 \ll Q^2$ and  $M^2 \ll Q^2$, with  $\bm P_{hT}$ being the component of the three-momentum $\bm P_h$ transverse to $\bm {q}$ and $M$ the nucleon mass.

The structure function $F_{UU,T}$ can be conveniently expressed in terms of Fourier transforms of the distribution  $f_1^a\big(x, \bm{k}_\T^2;Q^2\big)$ and fragmentation function  $D_1^{a\smarrow h}\big(z, \bm{P}_{\T}^2; Q^2 \big)$ for a quark with flavor $a$ and electric charge $e_a$ in units of the proton charge, which are defined by
\begin{align} 
\tilde{f}_1^a\big(x, \bT^2;Q^2\big) &=
\int_0^{\infty} d |\bm{k}_\T| 
                |\bm{k}_\T|J_0\big(\bT |\bm{k}_\T|\big) 
       f_1^a\big(x, \bm{k}_\T^2;Q^2\big)\,,\\
\tilde{D}_1^{a\smarrow h}\big(z, \bT^2; Q^2 \big) &=
\int_0^{\infty} \frac{d |\bm{P}_{\T}|}{z^2} |\bm{P}_{\T}| 
                                             J_0\big(\bT |\bm{P}_{\T}|/z\big)
       D_1^{a\smarrow h}\big(z, \bm{P}_{\T}^2; Q^2 \big)~.
\end{align}  
To leading order (LO) in perturbative QCD and in the region $\bm{P}_{hT}^2 \ll Q^2$, $F_{UU,T}$ reads
\begin{align}
\label{e:SIDISkTFF}
   F_{UU,T}(x,z, \bm{P}_{h \Tperp}^2, Q^2) &\approx 2\pi \sum_a e_a^2 x 
       \int_0^{\infty} {d \bT} \bT J_0\big(\bT |\bm{P}_{hT}|/z\big)
      \tilde{f}_1^a\big(x, \bT^2;Q^2\big) \tilde{D}_1^{a\smarrow h}\big(z, \bT^2;
      Q^2 \big)~. 
\end{align} 

Similarly, the  cross sections for the DY and $Z$ production processes, $h_A(P_A) + h_B(P_B) \to \ell^+(l) + \ell^-(l^\prime) + X$, can be written as
\begin{align}
\label{e:dsigma_gZ}
\frac{d\sigma}{dQ^2\, dq_T^2\,d\eta} &= \sigma_0^{\gamma,Z}
\bigg(F_{UU}^1 + \frac{1}{2} F_{UU}^2\bigg)\,,
\end{align} 
where $q$ and $\eta$ are respectively the four-momentum and rapidity (w.r.t.~the direction defined by $P_A$) of the virtual photon or $Z$ boson exchanged in the reaction, while $Q^2 = q^2$. The explicit expressions  for the elementary cross sections $\sigma_0^{\gamma,Z}$ can be found  in Ref.~\cite{Bacchetta:2017gcc}. At LO, in the kinematic limit $q_T^2 \ll Q^2$, $F_{UU}^2 \approx 0$ and 
\begin{align}
\label{e:DYkTFF}
   F_{UU}^1(x_A,x_B, \bm{q}_T^2, Q^2) &\approx
 2\pi \sum_a {\cal H}_{UU}^{1 a} \, \int_0^{\infty} d \bT \bT\, J_0\big( \bT |\bm{q}_T|\big)\ 
      \tilde{f}_1^a\big(x_A, \bT^2;Q^2\big) \   \tilde{f}_1^{\bar{a}}\big(x_B, \bT^2;Q^2 \big)  \, ,
\end{align} 
where  ${\cal H}_{UU, \gamma}^{1 a}(Q^2) \approx {e_a^2}/{N_c}$, ${\cal H}_{UU, Z}^{1 a}(Q^2) \approx {[ (I_{3a} - 2 e_{a} \sin^2 \theta_W)^2 + I_{3a}^2]}/{N_c} $, 
with $N_c$ being the number of colors, $\theta_W$ the Weinberg angle,  and $I_{3a}$ the weak isospin. The longitudinal momentum fractions $x_{A}$, $x_{B}$ can be expressed in terms of the rapidity $\eta$ by the relation  $x_{A/B} = {Q}/{\sqrt{s}}\,e^{\pm \eta}$.

\section{TMD evolution}
\label{sec:evol}

The Fourier transforms of the TMDs, evolved at the scale $Q^2$, at LO can be written as~\cite{Bacchetta:2017gcc,Collins:2011zzd}
\begin{align}   
\widetilde{f}_1^a (x,  \bT^2; Q^2) &= f_1^a (x ; \mu_b^2) 
\  e^{S (\mu_b^2, Q^2)} \  e^{g_K(\bT) \ln (Q^2 / Q_0^2)} \  \widetilde{f}_{1 {\rm NP}}^a (x, \bT^2) \ ,
\label{e:TMDevol1b} \\
\widetilde{D}_1^{a\to h} (z, \bT^2; Q^2) &= D_1^{a\to h} (z; \mu_b^2) \  e^{S (\mu_b^2, Q^2)} \  e^{g_K( \bT) \ln (Q^2 / Q_0^2)} \  \widetilde{D}_{1 {\rm NP}}^{a\to h} (z, \bT^2) \  ,
\label{e:TMDevol2b}
\end{align}
where $f_1^a (x ; \mu_b^2) $ and $D_1^{a\to h} (z; \mu_b^2) $ are the common  collinear distribution and fragmentation functions, evaluated at an initial scale $\mu_b$. The evolution is driven by the Sudakov exponent\footnote{We point out that a factor $1/2$ is missing in the definition of the Sudakov factor  in (2.3) of \cite{Bacchetta:2017gcc}.}
\begin{equation} 
S(\mu_b^2,Q^2)=- \frac{1}{2}\int_{\mu_b^2}^{Q^2}{d\mu^2\over \mu^2}
\bigg[A\Big(\alpha_S(\mu^2)\Big)\ln\bigg({Q^2\over \mu^2}\bigg) 
+ B\Big(\alpha_S(\mu^2)\Big) \bigg] \ ,
\label{e:Sudakov} 
\end{equation} 
where, at the next-to-leading logarithmic (NLL) accuracy, 
\begin{align}
A &= C_F\bigg(\frac{\alpha_S}{\pi} \bigg) +  \frac{1}{2}\,C_F
\bigg(\frac{\alpha_S}{\pi} \bigg)^2 \bigg[
C_A \bigg( \frac{67}{18} - \frac{\pi^2}{6} \bigg)
- \frac{5}{9} N_f \bigg],
&
B &= - \frac{3}{2}C_F
\bigg(\frac{\alpha_S}{\pi} \bigg)\,,
\end{align} 
with $C_F = (N^2_c-1)/2N_c$, $C_A=N_c$  and $N_f$ the number of active flavors. 

The  initial scale is chosen to be $\mu_b = {2 e^{-\gamma_E}}/{\bar{\bb}_{\ast}} $, where $\gamma_E$ is the Euler constant and
\begin{align} 
\bar{\bb}_{\ast} &\equiv \bar{\bb}_{\ast}(\bT;\bb_{\rm min},\bb_{\rm max}) = \bb_{\rm max} \Bigg(\frac{1-e^{- \bT^4 / \bb_{\rm max}^4} }{1-e^{- \bT^4 / \bb_{\rm min}^4}} \Bigg)^{1/4} .
\label{e:b*}
\end{align}  
When performing the integrals in (\ref{e:SIDISkTFF}) and (\ref{e:DYkTFF}), the variable $\bar{\bb}_{\ast}$ replaces $\bT$ in the perturbative parts of the TMDs defined in~\eqref{e:TMDevol1b}-\eqref{e:TMDevol2b}, because at large $\bT$ perturbative calculations are not reliable any more. The
  $\bar{\bb}_{\ast}$ is assumed to saturate on the maximum value $\bb_{\rm max}$, as suggested also in Refs.~\cite{Collins:2011zzd,Aybat:2011zv}. Numerical values are taken to be
\begin{align}
\bb_{\rm max} &= 2 e^{-\gamma_E}  \text{  GeV}^{-1} \approx 1.123 \text{  GeV}^{-1}, 
&
\bb_{\rm min} &= 2 e^{-\gamma_E}/Q \ .
\label{e:bminmax}
\end{align} 
In this way,  the scale $\mu_b$ is constrained to be always between 1 GeV and $Q$, such that the collinear distributions and fragmentation functions are never computed at a scale lower than 1 GeV.  Furthermore, in the definition of the perturbative Sudakov exponent~(\ref{e:Sudakov}), the lower integration bound is always smaller than the upper one. When $Q= Q_0=1$ GeV, $\bb_{\rm min}  = \bb_{\rm max}$ and no evolution effects are present: at this scale TMDs are simply given by the corresponding collinear function multiplied by a nonperturbative contribution.

In (\ref{e:TMDevol1b}) and (\ref{e:TMDevol2b}), the nonperturbative Sudakov factor is chosen to be $g_K (\bT) = - g_2 \bT^2 / 4$,
where $g_2$ is a free parameter. 

\begin{table}[t]
\small
  \centering
  \begin{tabular}{|c||c|c|c|c|c|c|}
\hline
\hline
TMD $f_1^a$ &  $g_1$ 
& $\alpha$ & $\sigma$ & & $\lambda$ &  
 \\ 
        & {[GeV$^2$]}                               &
       &      &  &{[GeV$^{-2}$]} & \\
\hline
All replicas &  $0.28\pm 0.06$ & $2.95\pm 0.05$ & $0.17\pm 0.02$ & 
                & $0.86\pm 0.78$ & 
\\
\hline
\hline
TMD $D_1^a$ &  $g_3$ &
$\beta$ & $\delta$ & $\gamma$ & $\lambda_F$ & $g_4$
 \\ 
        & {[GeV$^2$]} &            &        & &{[GeV$^{-2}$]} &{[GeV$^2$]}    \\
\hline
All replicas & $0.21\pm 0.02$ & $1.65\pm 0.49$ & $2.28\pm 0.46$ & $0.14\pm 0.07$ &
$5.50\pm 1.23$ & $0.13\pm 0.01$ \\
\hline
\hline
\end{tabular}
\caption{Best-fit values for the input parameters of the TMDs at the scale $Q=1$ GeV.}
\label{t:fl_ind_par_TMD}
\end{table}

The nonperturbative part of the distribution function in (\ref{e:TMDevol1b}) is parametrized as 
\begin{align}
\widetilde{f}_{1 {\rm NP}}^a (x, \bT^2) &= \frac{1}{2\pi}
        e^{-g_{1a} \frac{\bT^2}{4}}
        \bigg( 1 - \frac{\lambda  g_{1a}^{2}}{1+\lambda g_{1a}}  \frac{\bT^2}{4} \bigg)\  ,
\label{e:f1NP} 
\end{align}
where $\lambda$ is a free parameter and the Gaussian width $g_1$ is taken to be dependent on the longitudinal momentum fraction $x$ as follows,
\begin{equation} 
 g_1 (x) = N_1 \;  
\frac{(1-x)^{\alpha} \  x^{\sigma} }{ (1 - \hat{x})^{\alpha} \  \hat{x}^{\sigma} } \, ,
\label{e:kT2_kin}
\end{equation}
with $\hat{x} = 0.1$, while $\alpha, \, \sigma$,  $N_1 \equiv  g_1 (\hat{x})$ are also free parameters.
Similarly, the nonperturbative fragmentation function in (\ref{e:TMDevol2b}) is given by  the combination of two Gaussians,
\begin{align}
\widetilde{D}_{1 {\rm NP}}^{a \to h} (z, \bT^2) &= 
    \frac{ g_{3 a\to h} \   e^{-g_{3 a\to h} \frac{\bT^2}{4 z^2}}
        + \big(\lambda_F/z^2\big)    g_{4 a\to h}^{2}
    \left(1 - g_{4 a\to h} \frac{\bT^2}{4 z^2} \right)
         \  e^{- g_{4 a\to h} \frac{\bT^2}{4z^2}}}
     {2 \pi z^2 \Big(g_{3 a\to h} + \big(\lambda_F/z^2\big)    g_{4 a\to h}^{2}\Big)} \,,
\label{e:D1NP}
\end{align} 
with $\lambda_F$ has to be fitted to the data. The $z$-dependent Gaussian widths $g_{3,4}$ are parametrized according to 
\begin{align}  
g_{3,4} (z) = N_{3,4} \  
               \frac{ (z^{\beta} + \delta)\ (1-z)^{\gamma} }{ (\hat{z}^{\beta} + \delta)\   (1 - \hat{z})^{\gamma} } \, ,
 \label{e:PT2_kin}
 \end{align}
where $\hat{z} = 0.5$, and  $\beta, \, \gamma, \, \delta $, $N_{3,4} \equiv g_{3,4} (\hat{z})$  are the free parameters.  Hence, the average transverse momentum squared of the distribution and fragmentation functions at the initial scale $Q = Q_0$ read, respectively, 
\begin{align}
\big \langle \bm{k}_{\perp}^2 \big \rangle (x) &= \frac{g_1(x) + 2 \lambda g_1^2(x)}
{1+ \lambda g_1(x)},
&
\big \langle \bm{P}_{\perp}^2 \big \rangle (z) &= \frac{g_3^2(z) + 2 \lambda_F
  g_4^3(z)}{g_3(z) + \lambda_F g_4^2(z)}~.
\label{e:transmom2}
\end{align}

\begin{figure}[t]
\begin{center}
\hspace*{-1.7cm}
\includegraphics[trim = 0cm 10.2cm 0cm 4.6cm , clip, width=1.25\textwidth]{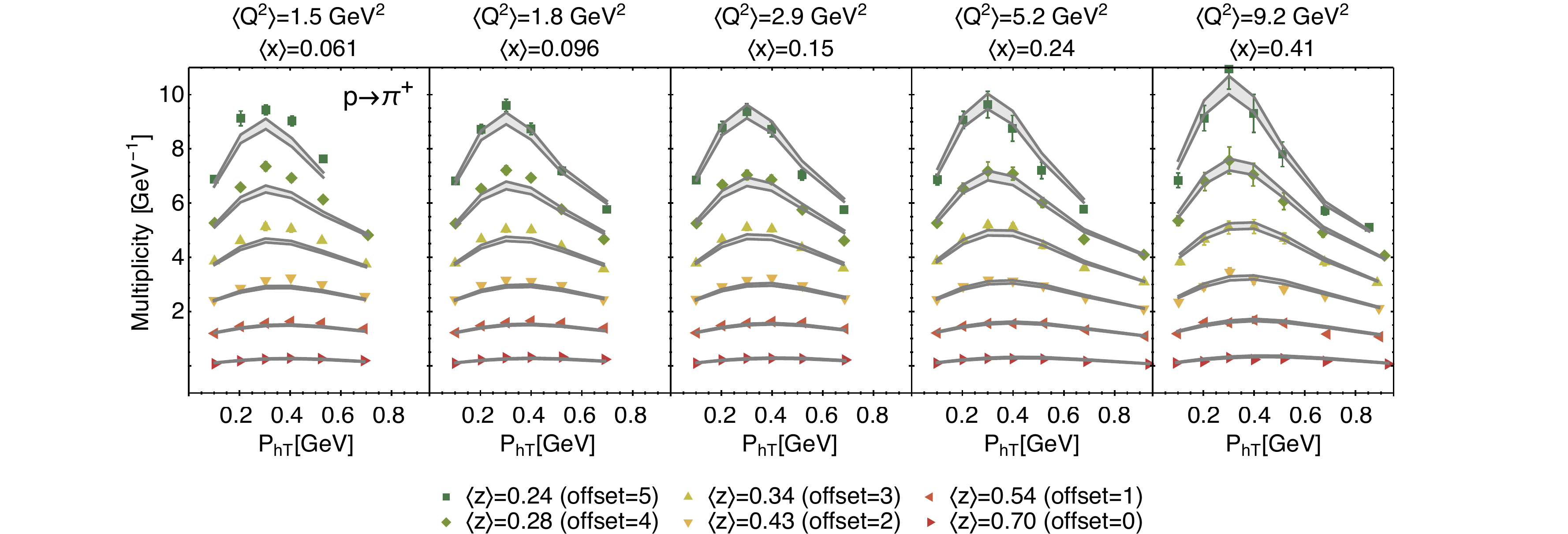}
\vspace{-0.7cm}
\end{center}
\caption{Multiplicities for the process $ep\to e \pi^+X$ at \hermes, shown as a function of the transverse momentum of the detected pion $P_{hT}$ in different bins of $\langle x \rangle$ , $\langle z \rangle$, $\langle Q^2 \rangle$.  For clarity, each $\langle z \rangle$  bin has been shifted by an offset as indicated in the legend.} 
\label{f:H_pions}
\end{figure}
\begin{figure}[t]
\begin{center}
\includegraphics[trim = 5cm 0cm 4cm 1cm , clip,angle = -90, width=1.\textwidth]{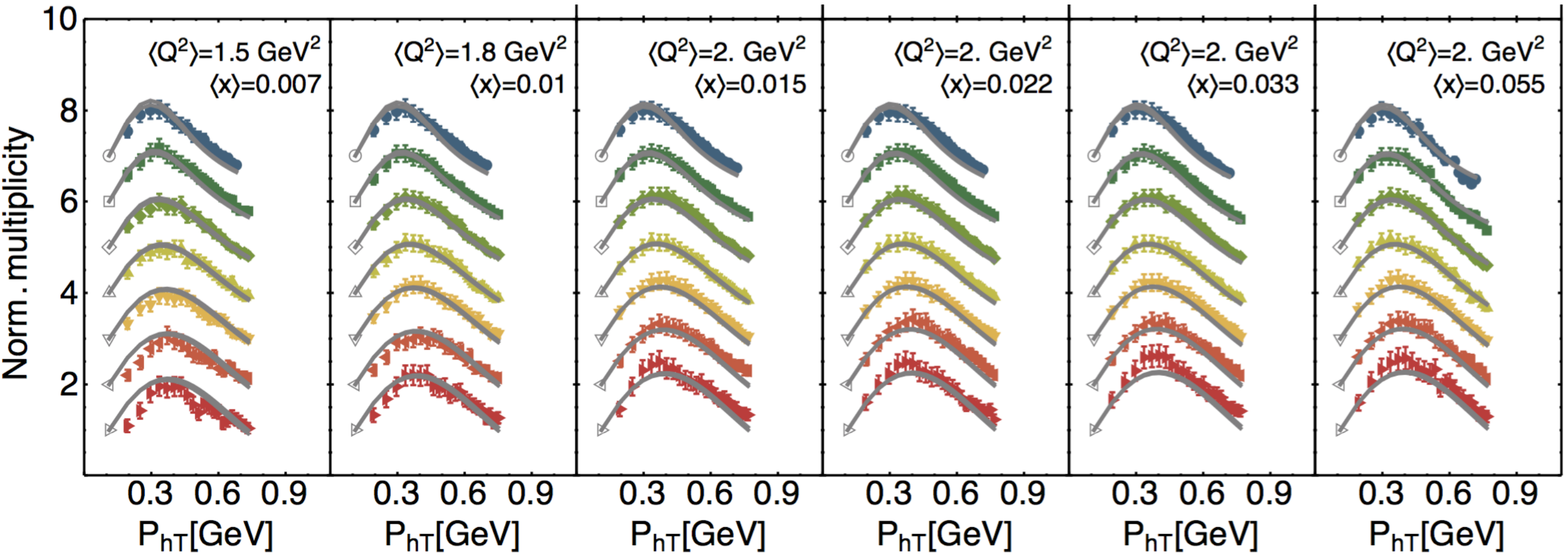}\\
\vspace{-0.8cm}\hspace*{-3cm}
\includegraphics[trim = 7cm 15.9cm 7cm 12cm , clip, width=0.18\textwidth]{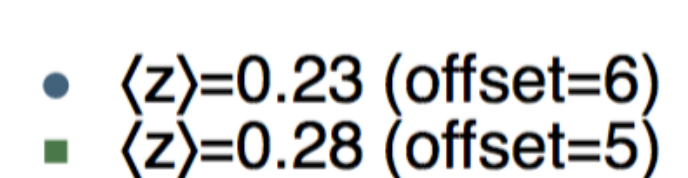}\hspace*{0.5cm}
\includegraphics[trim = 7cm 14.5cm 7cm 13.85cm , clip, width=0.18\textwidth]{offset-compass}\hspace*{0.5cm}\\
\vspace{-0.6cm} \hspace*{7.cm}
\includegraphics[trim = 7cm 12cm 7cm 15.2cm , clip, width=0.18\textwidth]{offset-compass}
\vspace{-0.6cm}
\end{center}
\caption{Multiplicities for the process $eD \to e h^+X$  at \compass, shown as a function of the transverse momentum of the detected hadron $P_{hT}$ at different $\langle x \rangle$ , $\langle z \rangle$, $\langle Q^2 \rangle$ bins. For each value of $\langle z \rangle$, multiplicities are normalized to the first bin in $P_{hT}$.  For clarity, each $\langle z \rangle$  bin has been shifted by an offset as indicated in the legend. } 
\label{f:C_pions}
\end{figure}

\section{Numerical results}
\label{sec:res}

The following data sets are included in our analysis:  SIDIS events off proton~\cite{Airapetian:2012ki} and deuteron targets~\cite{Airapetian:2012ki,Adolph:2013stb}, DY bins at low energy~\cite{Ito:1980ev,Moreno:1990sf} and $Z$ boson production at the Tevatron~\cite{Affolder:1999jh,Abbott:1999wk,Aaltonen:2012fi,Abazov:2007ac}. 

Concerning the kinematical cuts applied, we impose $Q^2 > 1.4$ GeV$^2$ and consider only the small transverse momentum region by selecting the maximum value of the transverse momenta measured in each process on the basis of phenomenological  considerations~\cite{Bacchetta:2017gcc}. In this way, it is possible to identify two different momentum scales, as required by TMD factorization. Furthermore, the current fragmentation region in SIDIS is isolated by implementing the cut $0.2 < z < 0.7$. After this selection, the total number of analyzed experimental bins is 8059. 

\begin{figure}[t]
\centering
\includegraphics[trim = 0cm 0.3cm 0cm 1.9cm , clip,width=1.\textwidth]{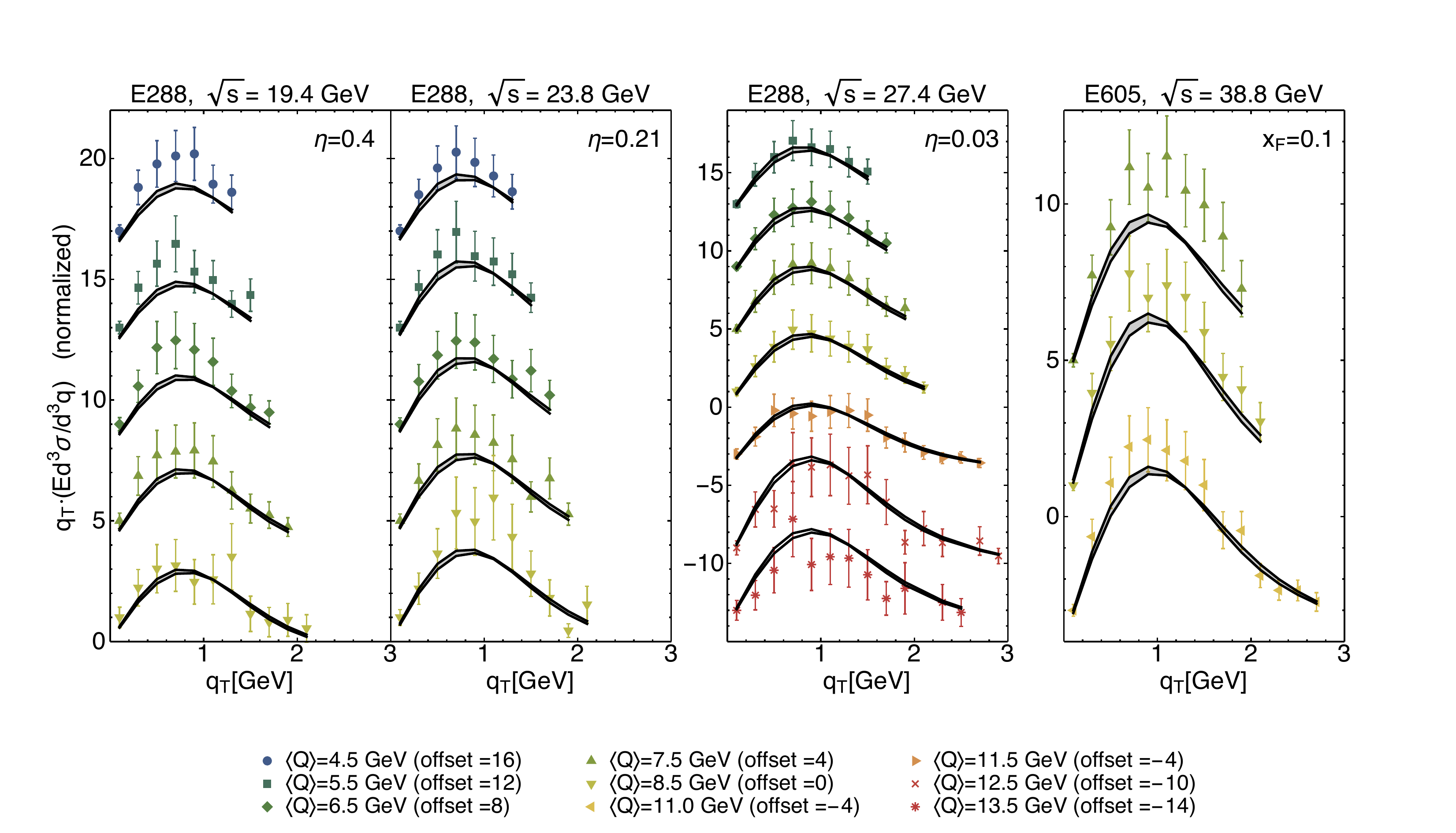}
\caption{DY cross section for several experiments, shown as a function of $q_T$ at different values of $\sqrt{s}$, $\eta$ and  $\langle Q \rangle$. For clarity, each $\langle Q \rangle$ bin has been normalized in such a way that the first data point is always one, and then shifted by an offset as indicated in the legend.}
\label{f:DY_panel}
\end{figure}
\begin{figure}[t]
\begin{center}
\includegraphics[trim = 0cm 1.3cm 0cm 2cm , clip,width=0.9\textwidth]{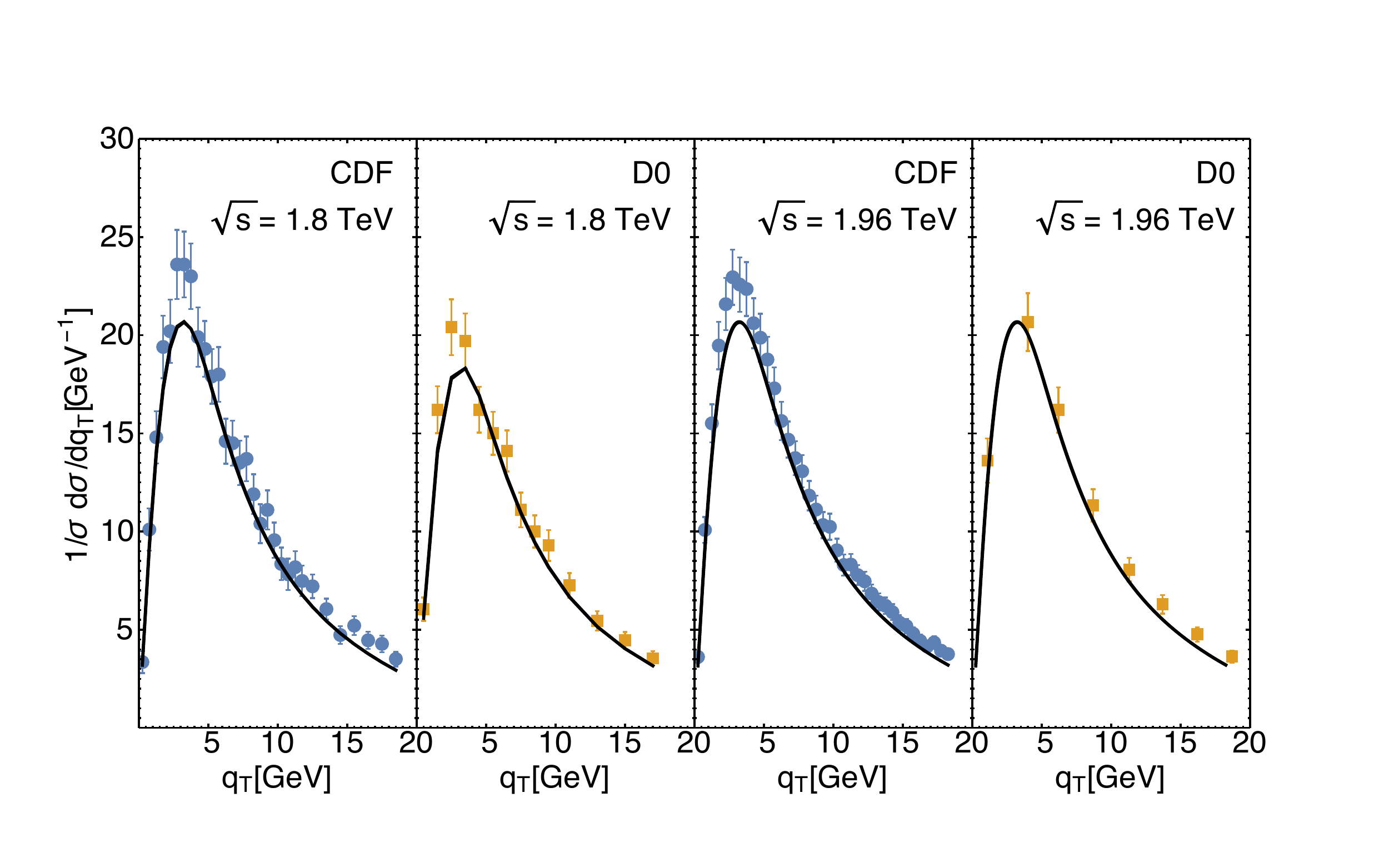}
\end{center}
\caption{Cross section for the process $p\bar p \to Z\, X$ for the CDF and D0 experiments at the Tevatron, shown as a function of  the transverse momentum $q_T$ of the $Z$ boson.} 
\label{f:Z_qT}
\end{figure}
For the collinear distributions $f_1^a$ in (\ref{e:TMDevol1b}) we adopt the GJR08FFnloE parametrization~\cite{Gluck:2007ck}, while 
for the collinear fragmentation functions $D_1^{a\to h}$ in (\ref{e:TMDevol2b}) the DSS14 NLO set for pions~\cite{deFlorian:2014xna} and the DSS07 NLO set for kaons~\cite{deFlorian:2007aj} are used.
  
The ten nonperturbative parameters of the TMDs $f_1^a$ and $D_1^{a\to h}$ at $Q = 1$  GeV are fitted to the data using a replica methodology~\cite{Bacchetta:2017gcc,Signori:2013mda}. The outcome is presented in Table~\ref{t:fl_ind_par_TMD}, where we give the avarage values of the parameters over the full set of 200 replicas, with a standard deviation based on the  68\% confidence level (C.L.)~\cite{Bacchetta:2017gcc}. In addition, we find that the nonperturbative parameter of the evolution, also fitted to the data,  is $g_2 =  (0.13 \pm 0.01)$ GeV$^2$. 

The average $\chi^2$/d.o.f $ = 1.55 \pm 0.05$ is rather good. This value can be further reduced to 1.02 without changing any free parameter, just by restricting the kinematical cuts such that it is possible to better identify the region where TMD factorization is supposed to hold.

For illustration, our resulting multiplicities for the SIDIS process $ep\to e \pi^+X$ at \hermes\ are presented in Fig.~\ref{f:H_pions} as a function of  the transverse momentum of the detected  pion $P_{hT}$, at different values of the $\langle x\rangle$, $\langle z \rangle $, $\langle Q^2\rangle$ bins. The bands are computed as the $68\%$ C.L.~envelope of the full sets of curves from all the replicas.
Similarly, in Fig.~\ref{f:C_pions} we show the multiplicities for positive hadron production off a deuteron target at \compass. Notice that, since \compass~measurements are affected by normalization errors~\cite{Adolph:2013stb},  we have fitted {\em normalized}  multiplicities. These are obtained by dividing the data in each bin in $(x,z,Q^2)$  by the data point with the lowest $P_{hT}$ in the bin, which is not  included in the analysis. 

In Figs.~\ref{f:DY_panel} and \ref{f:Z_qT} we compare, respectively,  the cross sections for the DY and $Z$ boson production processes with the measurements from different experiments. The results are differential in the transverse momentum of the exchanged virtual boson $q_T$. It is interesting to note that  the position of the peak moves from $q_T\sim 1$ GeV for DY events to $q_T\sim 7$ GeV for $Z$ boson production, as an effect of TMD evolution.

\begin{figure}[t]
\vspace{-0.2cm}
\centering
\subfloat[]{\includegraphics[width=0.45\textwidth]{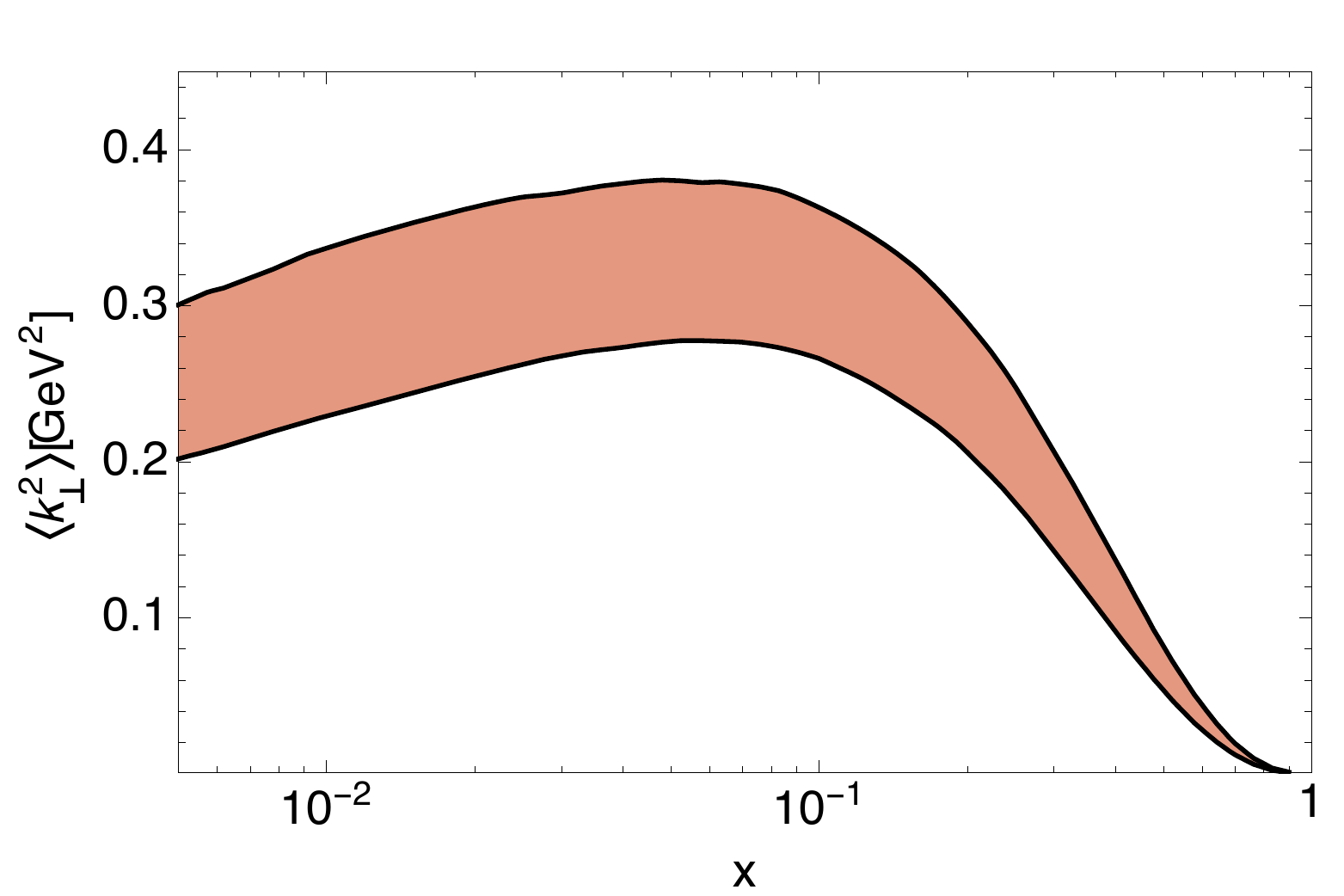}}
\hspace*{0.5cm}
\subfloat[]{\includegraphics[width=0.45\textwidth]{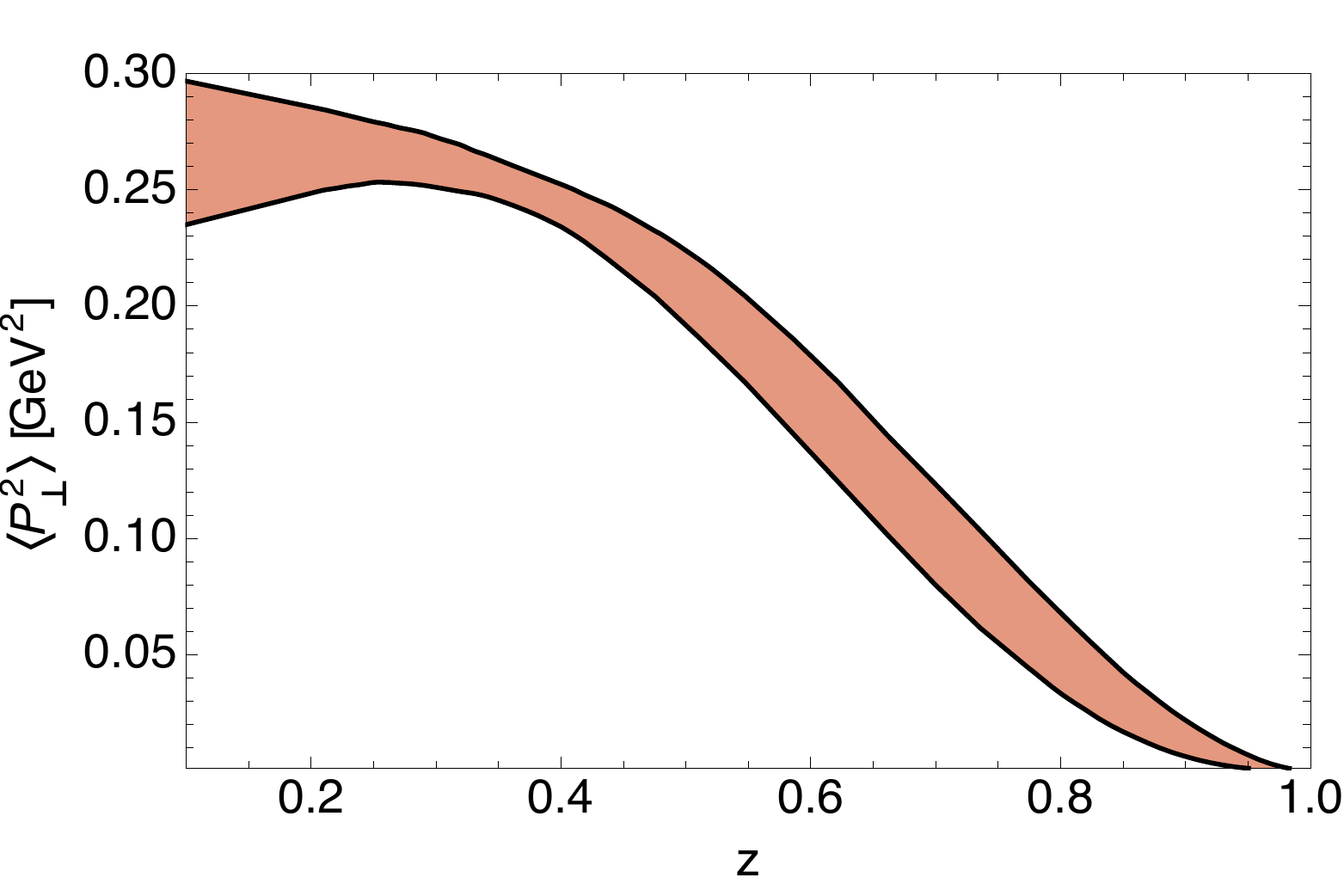}}
\caption{$\big \langle \bm{k}_{\T}^2 \big \rangle$ in the TMD distributions as a function of $x$
  (a), and $\big \langle \bm{P}_{\perp}^2 \big \rangle$ in the TMD fragmentations as a function of $z$ (b), both calculated at $Q^2= 1$ GeV$^2$. The fitted data cover the kinematic regions $ 5 \times 10^{-3}\lesssim x \lesssim  0.5$  and $0.2 \lesssim z \lesssim 0.7$. }
\label{f:avmomenta_68CL}  
\end{figure}

Finally, we comment on the  average transverse momentum squared obtained for the incoming parton, $\big \langle \bm{k}_{\T}^2 \big \rangle$, and the one acquired during the fragmentation process, $\big \langle \bm{P}_{\perp}^2 \big \rangle$. They are defined by~\eqref{e:transmom2} at the initial scale $Q = 1$ GeV, where the TMDs coincide with their nonperturbative input, in terms of the fit parameters.  Their dependence on the longitudinal momentum fractions, respectively $x$ and $z$, are presented in Fig.~\ref{f:avmomenta_68CL}.  

\begin{figure}[t]
\begin{center}
\includegraphics[width=0.60\textwidth]{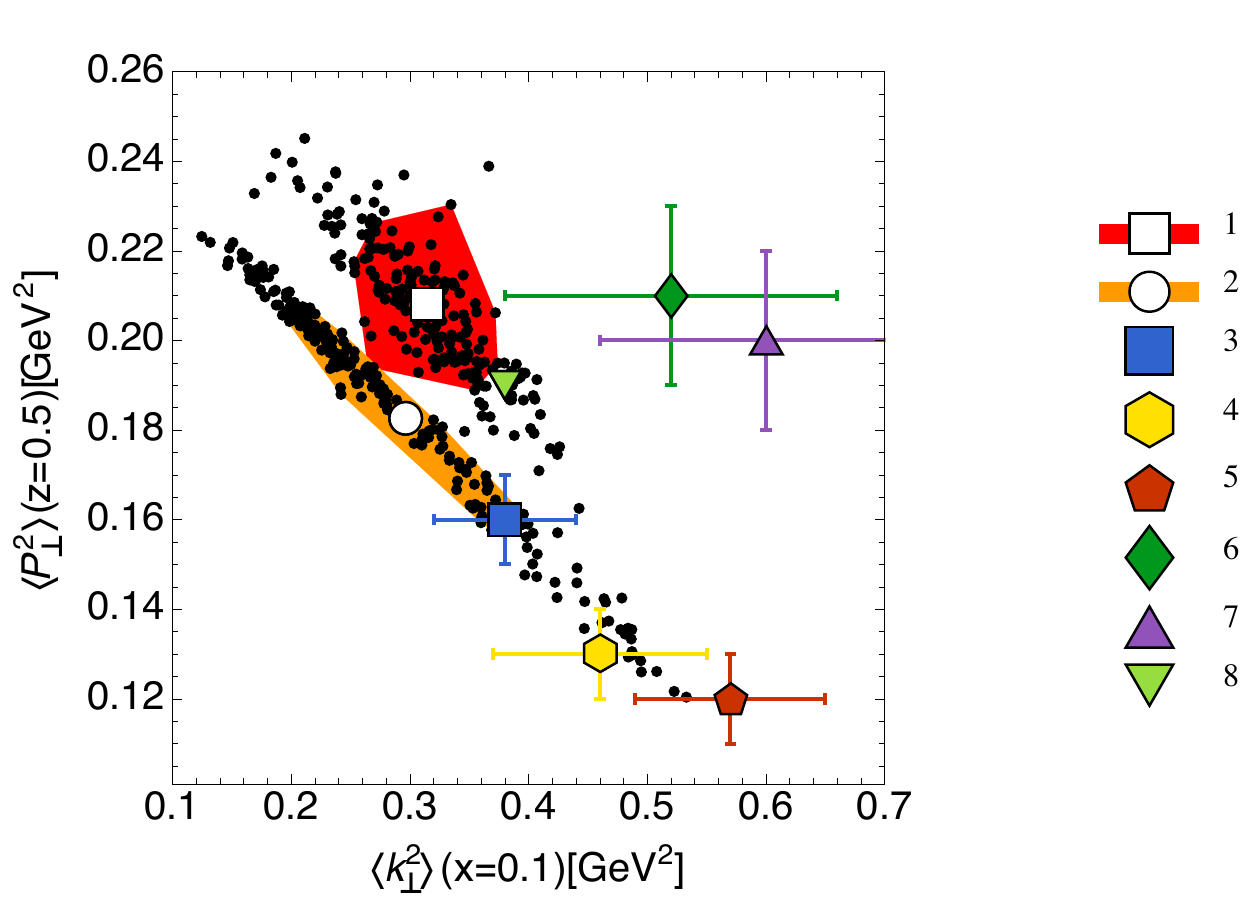}
\end{center}
\caption{Correlation between the transverse momenta in the TMD fragmentation functions, $\langle P_\perp^2
  \rangle(z=0.5)$, and in the TMD distribution functions, $\langle k_\perp^2 \rangle(x=0.1)$, according to  different phenomenological extractions: (1) average values obtained in the present analysis [shown together with the values from each replica (black dots) and
 the $68\%$ C.L.~area (red)]; (2) results from~\cite{Signori:2013mda},
(3) from \cite{Schweitzer:2010tt}, (4) from~\cite{Anselmino:2013lza} for
 \hermes\ data, (5) from~\cite{Anselmino:2013lza} for \hermes\ data at high $z$, (6) from~\cite{Anselmino:2013lza} for normalized \compass\ data, (7) from~\cite{Anselmino:2013lza} for normalized \compass\ data at high $z$, (8) from~\cite{Echevarria:2014xaa}.} 
\label{f:kT2_vs_PT2}
\end{figure}

In Fig.~\ref{f:kT2_vs_PT2} we show the values of $\big \langle \bm{k}_{\T}^2 \big \rangle$  at $x=0.1$ in the horizontal axis,  and of $\big \langle \bm{P}_{\perp}^2 \big \rangle$ at  $z=0.5$ in the vertical axis. The white square (label 1) denotes the average values of these two quantities obtained in the present analysis,  both calculated at the same scale $Q=1$ GeV, while each black dot around the white square is the outcome of one replica. The red region represents the envelope of the $68\%$ of the replicas that are closest to the average value. In the figure, our results are compared with other extractions. In particular, the white circle and the orange region around it (label 2) refer to the flavor-independent scenario in~\cite{Signori:2013mda}, which was obtained neglecting TMD evolution and fitting only  SIDIS data from the \hermes\ experiment at an average $\langle Q^2 \rangle= 2.4$ GeV$^2$. The older analysis shows a strong anticorrelation between the transverse momenta, which is reduced in the present study. This effect is due to the inclusion of Drell--Yan and $Z$ production data, that add new physical information about TMD parton distributions without the influence of fragmentation functions. In the two analyses, the average values of $\big \langle \bm{k}_{\T}^2 \big \rangle$ at  $x=0.1$ are similar and compatible within error bands, while 
the present values of $\big \langle \bm{P}_{\perp}^2 \big \rangle$ at $z=0.5$  turn out to be larger than in the older study,  mainly as a consequence of the \compass\ data. In general, the $x$ and $z$ dependence of the transverse momentum squared are also different in the two extractions. 

\section{Conclusions}
\label{sec:conc}

We have performed the first extraction of unpolarized quark TMDs from a simultaneous fit of SIDIS, Drell-Yan and $Z$ boson production data, taken in several experiments at different energies.  Our analysis is restricted to the the small transverse momentum region and implements TMD evolution effects up to the NLL accuracy. Our $\chi^2$/d.o.f.~$ = 1.55 \pm 0.05$ can be further improved to 1.02 without changing the parameters, only by restricting the kinematic cuts.  It turns out  that most of the discrepancies with the measurements come from the normalization rather than from the transverse momentum shape. Hopefully, this tension could be reduced by a more precise treatment from the perturbative point view~\cite{Scimemi:2017etj}.  

In future studies, we plan to improve our analysis by exploring different functional forms for all the nonperturbative quantities, possibly including a  flavor dependence of the intrinsic transverse momenta. Furthermore, one should try to properly match the description at low transverse momentum with the fixed-order results at high transverse momentum, obtained within the framework of collinear factorization.

\begin{acknowledgments}
This work is supported by the European Research Council (ERC) under the European Union's Horizon 2020 research and innovation program (grant agreement No. 647981, 3DSPIN). AS acknowledges support from U.S. Department of Energy contract DE-AC05-06OR23177, under which Jefferson Science Associates, LLC, manages and operates Jefferson Lab. Preprint number: JLAB-THY-18-2628.
\end{acknowledgments}

\end{document}